\documentclass[preprint,aps,prd,groupedaddress,showpacs]
              {revtex4}%
\usepackage{graphicx}
\usepackage{amsmath}
\usepackage{bm}

\begin{document}

\preprint{\begin{tabular}{l}
          ANL-HEP-PR-03-027\\
          FERMILAB-PUB-03/067-T\end{tabular}}


\bigskip

\title{Squark Mixing in Electron-Positron Reactions}

\author{Edmond L.\ Berger}
\email[e-mail: ]{berger@anl.gov}
\affiliation{High Energy Physics Division, 
Argonne National Laboratory, Argonne, IL 60439}
\author{Jungil Lee}
\email[e-mail: ]{jungil@hep.anl.gov}
\affiliation{High Energy Physics Division, 
Argonne National Laboratory, Argonne, IL 60439}
\author{Tim M.\ P.\ Tait}
\email[e-mail: ]{tait@fnal.gov}
\affiliation{Fermi National Accelerator Laboratory,
P.O. Box 500, Batavia, IL 60510} 

\date{\today}
\begin{abstract}

Squark mixing plays a large role in the phenomenology of the minimal 
supersymmetric standard model, determining the mass of the lightest Higgs
boson and the electroweak interactions of the squarks themselves.  We
examine how mixing may be investigated in high energy $e^+ e^-$ reactions,
both at LEP-II and the proposed linear collider.  In particular, off-diagonal
production of one lighter and one heavier squark allows one to measure
the squark mixing angle, and would allow one to test the mass relations for 
the light Higgs boson.
In some cases off-diagonal production may provide the best prospects 
to discover supersymmetry.
In the context of the light bottom squark scenario, we show that existing data 
from LEP-II should show definitive evidence for the heavier bottom squark 
provided that its mass $m_{\tilde{b}_2} \le 120$~GeV.  
\end{abstract}



\maketitle


\section{Introduction
	\label{intro}}

Scalar quarks, the supersymmetric partners of ordinary colored fermions, are
an important ingredient in any theory which combines the standard model (SM) with
supersymmetry (SUSY).  These scalar partners of the quarks play a key role 
in the softening of quadratic divergences in the Higgs boson mass
parameter induced by loops of quarks.  This reduced sensitivity to the 
ultra-violet physics is the single most attractive feature of the 
minimal supersymmetric standard model (MSSM).  Phenomenologically, scalar
quarks (squarks) provide interesting signatures at colliders.  They are color 
triplets,
and so may be produced copiously in hadronic reactions.  They also
couple electroweakly, and thus may have spectacular decays involving 
leptons and weak bosons.  In fact, as quarks in the SM are the fields which
`bridge' the strong and electroweak interactions, so squarks are the bridge
between the supersymmetric analogues of the gauge sectors.
As spin-0 fields, the angular distributions of their
production and decay are often quite distinct from their spin-1/2 counterparts.

Each massive quark of the SM results from the marriage (through electroweak
symmetry-breaking) of two Weyl fermions with very different electroweak 
characteristics.  Each Dirac fermion is accompanied by two complex
scalars, also with distinct electroweak interactions.  The supersymmetric 
versions of the Yukawa interactions between the Higgs boson and the fermions mix
this pair of scalars, leading to two mass eigenstates denoted $\tilde{q}_1$ 
and $\tilde{q}_2$.  The amount of mixing, represented in terms of 
a mixing angle $\theta_q$, is typically proportional
to the associated fermion's mass, and the mixing is thus presumably
largest for the third generation squarks.  

Squark mixing plays an interesting role in the phenomenology of the MSSM.
Perhaps foremost is the fact that the lightest MSSM Higgs boson, the one
primarily responsible for electroweak symmetry-breaking, has a mass at tree
level which is less than the $Z$ boson mass, considerably smaller than 
114.5 GeV, the experimental bound from the CERN Large Electron Positron 
(LEP) facility~\cite{lephiggs}.  
This expectation would exclude the
MSSM if it were not for the fact that the Higgs boson mass receives large
radiative corrections which can lift the mass beyond the reach of the
exclusion limits, to masses as large as about 135 GeV.  These necessary 
corrections are provided largely by the top squarks and are enhanced by the
squark mixing~\cite{Carena:1995wu},
\begin{eqnarray}
m_H & \simeq & M_Z^2 + \frac{3 g^2 m_t^4}{8 \pi^2 m_W^2} \left[
\log \left( \frac{M_S^2}{m_t^2} \right) 
+ x_t^2 \left( 1 - \frac{x_t^2}{12} \right) \right] .
\end{eqnarray}
In this expression, $m_t$ is the top quark mass, $M_S$ is a common scale of 
SUSY-breaking which describes the overall
magnitude of the squark masses, and 
$x_t = (A_t - \mu \cot \beta) / M_S$ characterizes the off-diagonal entries
in the squark mass matrix (and thus the amount of mixing; 
$x_t \propto \sin \theta_t \cos \theta_t$),
with the largest Higgs boson masses realized for $x_t \sim \sqrt{6}$.
Once the lightest Higgs boson and top squarks are discovered, a key
test that the Higgs boson properties follow the MSSM paradigm will require
careful measurement of the top squark masses and their mixing 
angle.  When this dominant correction is understood, one begins to
indirectly probe the remainder of the MSSM through the sub-dominant
one-loop and dominant two-loop corrections to the Higgs boson 
mass~\cite{Heinemeyer:1998jw}.

The mixing angle determines the couplings of the squark mass 
eigenstates 
to the $W$ and $Z$ bosons.  It can be determined efficiently 
if these couplings are measured.  For instance, the rate of the 
reaction $e^+ e^- \rightarrow \widetilde{q}_1^* \widetilde{q}_2$, production 
of one lighter and one heavier squark, is proportional to the amount
of mixing, and a measurement of the rate is an excellent way to 
establish the value of the mixing
angle.  Off-diagonal production provides an interesting complement to 
other proposals~\cite{Bartl:2000kw} to measure the mixing angle at an 
electron-positron 
linear collider~\cite{lcstudies}.  These measurements typically 
involve the cross section for top squark $\tilde t_1 \tilde t^*_1$ pair 
production with
various combinations of polarized $e^+$ and $e^-$ beams.  Off-diagonal 
production provides an important cross-check of these other methods,
and it does not rely on any particular polarization of the incoming beams.
Even if the heavier squark mass is larger than half of the 
collider energy, precluding pair production of the heavier squarks,
this method can still succeed and also allow one to measure the mass of
the heavier squark.  

This line of reasoning is indicative of a general feature of squark 
phenomenology.  While hadron colliders generally
offer the best prospects to {\em discover} squarks, due to copious
production through the strong interactions, it is somewhat problematic to
determine parameters in hadronic reactions.  In hadron collisions, the
primary means to study the electroweak properties of squarks is to examine 
squark decays.  This task is challenging for a number of reasons.  In many
popular models of SUSY-breaking~\cite{Abel:2000vs}, the squark decay is
dominated by a single channel (often into a quark and the lightest 
neutralino) dictated purely by kinematics, and not by couplings.  The 
branching ratios of
various decay modes are relatively insensitive to the 
squark mixing angle.  Furthermore, the squark decay width, the quantity
that actually depends on the coupling strength, is in almost all cases
well below the experimental resolution and thus unmeasurable at hadron
colliders.

This discussion illustrates the importance of squark mixing on 
the determination of bounds on squark masses from data at 
lepton colliders.  A particular example is furnished by the scenario of 
light bottom squarks and light gluinos proposed in 
Ref.~\cite{Berger:2000mp}.  In this scenario, the excess rate of bottom quark
production at hadron colliders is explained by postulating a tree-level
contribution from production of light gluinos which decay into 
a bottom quark and a bottom squark.  
Data indicate that the lifetime of the hypothesized light bottom squark 
must be less than 1 nanosecond~\cite{Janot:2003cr}; in typical collider detectors, 
it does not have a significant missing energy signature nor does it 
produce tracks characteristic of heavy long-lived objects.  
The light bottom squark $\tilde{b}_1$ is assumed to decay hadronically, 
via R-parity violation, without a visible flavor tag necessarily.  
Its signals are extremely difficult to extract from 
backgrounds~\cite{Berger:2000mp,Berger:2002kc}.  Furthermore, a light 
bottom squark evades LEP-I data if one uses the freedom to select a mixing 
angle which renders its coupling to the $Z$ boson tiny \cite{Carena:2000ka}.
This essential requirement implies a non-zero mixing angle
($\sin^2 \theta_b \sim 1/6$ ), and, therefore, the off-diagonal 
$Z$-$\widetilde{b}_1$-$\widetilde{b}_2$ coupling must be non-zero.  Here 
$\tilde{b}_2$ denotes the heavier of two bottom squarks.   
One of the most promising (and potentially clear) signals of this scenario
is furnished by 
$e^+ e^- \rightarrow Z^* \rightarrow \widetilde{b}_1 \widetilde{b}_2$.
References to other work on the phenomenology of light bottom squarks and 
light gluinos may be found in 
Refs.~\cite{Berger:2002kc,Berger:2002vs}.  

In this article we consider off-diagonal squark production at $e^+ e^-$ 
colliders.  We compute tree-level cross sections and discuss likely decay
modes.  In Sec.~\ref{sec:squarkproperties}, we
review the squark mass matrix and electroweak interactions.  
In Sec.~\ref{sec:xsec} we compute the production cross sections for 
both bottom squarks 
and top squarks, showing the dependence on the masses and mixing angles.
In Sec.~\ref{sec:decays} we apply our results to the light gluino and bottom
squark scenario at LEP-II, estimating for the first time the discovery 
potential of the heavier bottom squark to be greater than 5 standard 
deviations ($5 \sigma$) provided
$m_{\tilde{b}_2} \leq 120$ GeV.  Alternately, the LEP-II data can exclude
masses smaller than 130 GeV, if no signal is observed.
We reserve Sec.~\ref{sec:conclusions} 
for conclusions.

\section{Squark Masses and Electroweak Interactions}
\label{sec:squarkproperties}

In this section we briefly review the squark mass matrices, mixing angles,
and electroweak interactions.
The squark mass matrices are 
\begin{eqnarray}
& &
\left( \widetilde{t}_L^* \widetilde{t}_R^* \right)
\left(   
\begin{array}{cc}
M_L^2 + m_t^2 + D_L & m_t ( A_t - \mu \cot \beta ) \\
m_t ( A_t - \mu \cot \beta ) & M_{t_R}^2 + m_t^2 + D_{R}
\end{array}
\right)
\left(   
\begin{array}{c}
\widetilde{t}_L^* \\
\widetilde{t}_R^*
\end{array}
\right)
\end{eqnarray}
for top squarks, and
\begin{eqnarray}
& &
\left( \widetilde{b}_L^* \widetilde{b}_R^* \right)
\left(   
\begin{array}{cc}
M_L^2 + m_b^2 + D_L & m_b ( A_b - \mu \tan \beta ) \\
m_b ( A_b - \mu \tan \beta ) & M_{b_R}^2 + m_b^2 + D_{R}
\end{array}
\right)
\left(   
\begin{array}{c}
\widetilde{b}_L^* \\
\widetilde{b}_R^*
\end{array}
\right)
\label{eq:squarkmass}
\end{eqnarray}
for bottom squarks.  
Parameters $M_L$ and $M_{q_R}$ are the SUSY-breaking masses for left- and 
right-handed squarks,
$A_t$ and $A_b$ are the (SUSY-breaking)
trilinear interactions with the Higgs field, $\mu$ is the
Higgsino mass parameter, $\tan \beta$ is the ratio of Higgs boson vacuum 
expectation
values (VEVs), $m_b$ and $m_t$ are the bottom and top quark masses, and 
and $D_L$ and $D_{R}$ are $D$-terms,
$D_L = m_Z^2 \cos 2 \beta (T_3 - Q_f \sin^2 \theta_W)$, and
$D_{R} = Q_f m_Z^2 \sin^2 \theta_W \cos 2 \beta$, with $Q_f$ the quark charge,
and $T_3$ its weak isospin.  For simplicity, we neglect the possibilities of
phases in the soft-breaking parameters; their inclusion is straight-forward.

Diagonalization of the matrices determines the squark mass 
eigenstates characterized by mixing angles $\theta_q$.   
The physical squarks are a mixture of the scalar partners of the
left- and right-chiral quarks.  The mass eigenstates
are two complex scalars ($\tilde q_1$ and $\tilde q_2$), expressed 
in terms of left-handed (L) and right-handed (R) squarks, 
$\tilde{q}_L$ and $\tilde{q}_R$, as 
\label{eq:mixing}
\begin{eqnarray}
|\tilde{q}_1\rangle = \sin\theta_{q}|\tilde{q}_L\rangle + 
                                \cos\theta_{q}|\tilde{q}_R\rangle , 
\nonumber \\
|\tilde{q}_2\rangle = \cos\theta_{q}|\tilde{q}_L\rangle - 
                                \sin\theta_{q}|\tilde{q}_R\rangle ,   
\end{eqnarray}
where our convention is that $\widetilde{q}_1$ is the lighter of the two
mass eigenstates.  These angles may be expressed as 
\begin{subequations}
\begin{eqnarray}
\sin 2 \theta_t & = & \frac{2 m_t (A_t - \mu \cot \beta)}
{m_{\tilde{t}_1}^2 - m_{\tilde{t}_2}^2} , \\
\sin 2 \theta_b & = & \frac{2 m_b (A_b - \mu \tan \beta)}
{m_{\tilde{b}_1}^2 - m_{\tilde{b}_2}^2} .
\end{eqnarray}
\end{subequations}
It is worth emphasizing that the mixing terms in Eq.~(\ref{eq:squarkmass})
are proportional to the quark masses, and thus the mixing is presumably
largest for the third generation squarks.  For this reason, we focus on
scalar top and bottom quarks in the discussion below.  

The electroweak interactions of the squarks are determined by the relative
admixture of left- and right-chiral squark in the mass eigenstate.  The
coupling to the $SU(2)_L$ gauge bosons is only through the left-chiral 
component, whereas coupling to the $U(1)_Y$ boson is non-zero for both.
After electroweak symmetry breaking (EWSB), the photon remains massless, 
and its gauge invariance is linearly
realized.  All squarks (of a given electric charge) couple equally 
to the photon with coupling
strength given by $Q_f e$.  The $Z$ boson couplings, on the other hand,
are sensitive to the mixing angles,
\begin{eqnarray}
g_{11} & = & \frac{e}{\sin \theta_W \cos \theta_W} 
\left[ T_3 \sin^2 \theta_q - Q_f \sin^2 \theta_W \right] ,
\nonumber \\
g_{12} & = & \frac{e}{\sin \theta_W \cos \theta_W} 
\left[ T_3 \sin \theta_q \cos \theta_q \right] ,
\nonumber \\
g_{22} & = & \frac{e}{\sin \theta_W \cos \theta_W} 
\left[ T_3 \cos^2 \theta_q - Q_f \sin^2 \theta_W \right] ,
\end{eqnarray}
where $g_{ij}$ refers to the coupling of the $Z$ boson with $\tilde{q}_i$
and $\tilde{q}_j$.  As mentioned above, the coupling to the 
lighter squarks may be tuned to vanish for 
$\sin^2 \theta_b \simeq 1/6$ (for bottom squarks) and
$\sin^2 \theta_t \simeq 1/3$ (for top squarks)
\cite{Carena:2000ka}, but in these limits,
the off-diagonal couplings, and the heavy-heavy couplings are non-zero.

The couplings to the $W$ boson are
\begin{eqnarray}
g^W_{11} & = & \frac{e}{\sqrt{2} \sin \theta_W} 
\sin \theta_t \sin \theta_b ;\nonumber \\
g^W_{12} & = & \frac{e}{\sqrt{2} \sin \theta_W} 
\sin \theta_t \cos \theta_b ;\nonumber \\
g^W_{21} & = & \frac{e}{\sqrt{2} \sin \theta_W} 
\cos \theta_t \sin \theta_b ;\nonumber \\
g^W_{22} & = & \frac{e}{\sqrt{2} \sin \theta_W} 
\cos \theta_t \cos \theta_b ; 
\end{eqnarray}
where now $g_{ij}$ refers to the coupling of $W$
to $\tilde{t}_i$ and $\tilde{b}_j$.  Squark couplings to quarks and
either charginos $\widetilde{\chi}^\pm$ or neutralinos $\widetilde{\chi}^0$ 
are straightforward, but somewhat more complicated by the mixing angles 
associated with the $\widetilde{\chi}^\pm$ and $\widetilde{\chi}^0$ mass 
eigenstates.

\section{Production Cross Sections}
\label{sec:xsec}

\begin{figure}
\includegraphics[height=2.5cm]{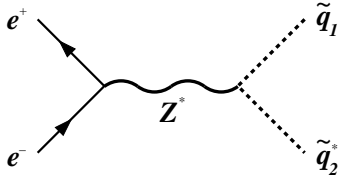}
\caption{Feynman diagram for the process 
$e^+ e^- \to \tilde{q}_1 {\tilde{q}}^*_2$.}
\label{fig:feynman}
\end{figure}

In this section, we examine the production of the off-diagonal pair of 
squarks $\tilde{q}_1$ and $\tilde{q}_2^*$ in the electron-positron 
annihilation process 
$e^+e^- \rightarrow \widetilde{q}_1 \widetilde{q}^*_2$, illustrated in 
Fig.~\ref{fig:feynman}.  The conjugate process
$e^+e^- \rightarrow \widetilde{q}^*_1 \widetilde{q}_2$
has the same cross section.  Each reaction has a single Feynman diagram
in which a $Z$-boson in exchanged in the $s$-channel.  The unbroken gauge
invariance of QED forbids the photon from contributing to off-diagonal
squark production.  For our purposes, it is enough to consider the tree-level
production rates.  Initial state radiation, and Yukawa and 
SUSY-QCD one-loop corrections are computed in Ref.~\cite{Drees:1990te}  
and can be typically as large
as $\pm 15\%$ for some regions of parameter space.

\subsection{$\bm{e^+e^-\to}$
     $\widetilde{\bm{q}}$$\bm{{}_1}$
     $\widetilde{\bm{q}}$$\bm{{}^*_2}$}

The amplitude for the process 
\begin{eqnarray}
&& e^-(k_1) + e^+(k_2)\to Z^* 
  \to \tilde{b}_1 (p_1,m_1) + \tilde{b}^*_2 (p_2,m_2)
\end{eqnarray}
is expressed as 
\begin{eqnarray}
\mathcal{M}&=&g_{12}
\bar{v}(k_2) \gamma_\mu (g_R P_R+g_L P_L)
\bar{u}(k_1)\frac{(p_1-p_2)^\mu}{s-M_Z^2} ,
\end{eqnarray}
where $P_L=(1-\gamma_5)/2$ and  $P_R=(1+\gamma_5)/2$. Specified are the 
four-momenta $k_1$ and $k_2$ 
of the incident $e^-$ and $e^+$, and $p_1$ and $p_2$ of the final 
squarks.  The lepton couplings to the $Z$ are 
\begin{subequations}
\begin{eqnarray}
g_L&=& \frac{e}{\sin\theta_W \cos\theta_W} 
\left( -\frac{1}{2} + \sin^2\theta_W \right);
\\
g_R&=&\frac{e}{\sin\theta_W \cos\theta_W} \left( \sin^2\theta_W \right).
\end{eqnarray}
\end{subequations}
Taking the absolute square of the amplitude, summing over final spins and 
colors, and averaging over the initial spins, we obtain the differential 
cross section 
\begin{eqnarray}
\frac{d\sigma}{d\cos\theta^*}&=&
\frac{r}{32\pi s} \overline{\sum} |\mathcal{M}|^2
 = \frac{3 g^2_{12} (g_L^2+g_R^2)}{128\pi s}
     \frac{r^3\sin^2\theta^*}{\left(1- M_Z^2/s \right)^2} ,
\label{eq:diffcross}
\end{eqnarray}
where
\begin{eqnarray}
r&=&\frac{2|\mathbf{p}^*|}{\sqrt{s}}
  =\frac{1}{s}\sqrt{ (s-m_1^2-m_2^2)^2 - 4 m_1^2 m_2^2 } .
\end{eqnarray}
Angle $\theta^*$ is the scattering angle of $\tilde{q}_1$ in the 
$e^+e^-$ center-of-mass frame, and ${\bf p^*}$ is its three-momentum.  
As expected for production of scalar particles, the energy dependence 
of the cross section is influenced by a $P$-wave threshold factor 
$\propto |\mathbf{p}^*|^3$, and the angular distribution varies as 
$\sin^2\theta^*$. 

The rate integrated over the scattering angle is 
\begin{eqnarray}
\sigma&=&
 \frac{g^2_{12} (g_L^2+g_R^2)}{32\pi s}
     \frac{r^3}{\left(1- M_Z^2/s \right)^2},
\end{eqnarray}
in which the dependence on the squark mixing angle $\theta_{q}$ is 
manifest in the proportionality to $\sin^2 2 \theta_{q}$ (in
the factor of $g_{12}^2$) in  Eq.~(\ref{eq:diffcross}).  
For reference, note that
production of a $\tilde{q}_1$ or a $\tilde{q}_2$ pair proceeds through both
photon and $Z$ exchange, and is not proportional to 
$\sin^2 2 \theta_{q}$, containing terms both sensitive and
insensitive to the mixing angle.

\subsection{Rates at a Linear Collider}

\begin{figure}
\includegraphics[height=11.0cm]{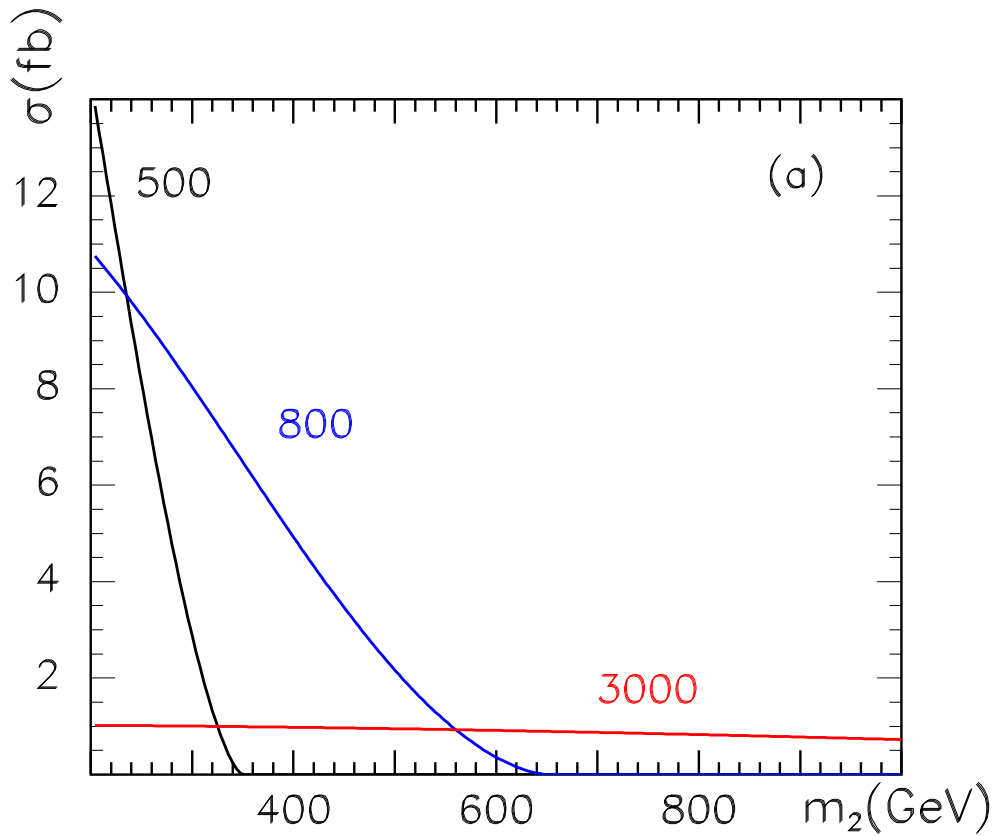}
\caption{Cross section for the process
$e^+ e^- \to \tilde{t}_1 {\tilde{t}}^*_2$ at center-of-mass 
energies 500, 800, and 3000 GeV, as a 
function of the mass of the heavier top squark and for mixing angle
$\sin^2 \theta_t = 1/2$. The mass of the lighter top squark has been fixed
to 150 GeV.}
\label{fig:stopxsec}
\end{figure}

As discussed above, a key measurement at a future linear collider would be
to verify the Higgs boson mass dependence on the supersymmetry-breaking
parameters.  This test would demonstrate that the MSSM is the 
effective theory at the weak scale, as opposed to some more general 
supersymmetric extension.  As in most of the MSSM parameter space, the
dominant corrections to the Higgs boson mass are from the top squarks, it is 
their
electroweak properties that are most relevant.  We envision (for illustrative
purposes) a situation in which the Higgs boson has been discovered at the 
Large Hadron Collider through some combination of production and decay
channels (see Ref.~\cite{atlas:1999fr}), and the squarks have been produced 
through
the strong interaction, dominantly $g g \rightarrow \tilde{t}_1^* \tilde{t}_1$
and $g g \rightarrow \tilde{t}_2^* \tilde{t}_2$.  The squark masses and
dominant decay channels are likely to be known, but the mixing angle (which
plays no role in the tree-level production through the strong force) must be
measured at a linear collider.

We consider, for reference, the light top squark to have a mass of 150 GeV,
somewhat above the Fermilab Run~I bounds~\cite{Demina:1999ty}, although the
bounds themselves are sensitive to the details of how the top squark decays.  
In Fig.~\ref{fig:stopxsec}, we show the cross sections 
for $e^+ e^- \rightarrow \tilde{t}_1^* \tilde{t}_2$ as a function of the
heavier top squark mass, for a reference mixing angle of 
$\sin^2 \theta_t = 1/2$.  We choose three center-of-mass energies:
$\sqrt{s} = 500$, $800$, and $3000$ GeV.  The rate is doubled if the 
charge conjugate process is included.  
We see that rates are typically of order a few femtobarns (fb) 
for top squark masses within the range allowed by kinematics. A linear
collider with hundreds of inverse fb of data could expect to produce 
(before cuts) hundreds of events, and the cross section could be measured
at the few per cent level, provided experimental efficiencies are not
extremely small and backgrounds not prohibitively large.  Such questions
must be answered in the context of specific top squark decay signatures 
and are not addressed in this work.

\section{Light Bottom Squarks and LEP-II}
\label{sec:decays}

As our second example, we consider bottom squark production and adopt 
parameters suggested in the light bottom squark scenario of 
Ref.~\cite{Berger:2000mp}.  This scenario postulates that the excess rate 
of bottom quark production at hadron colliders arises from pair production 
of gluinos with masses on the order of 15 GeV. The gluinos subsequently decay 
to bottom quarks and the light bottom squarks, with masses of order the bottom 
mass.
In order for such light scalar bottom quarks to be consistent with $Z$-pole
data, the light bottom squarks must decouple from the $Z$, implying a non-trivial
mixing angle $\sin^2 \theta_{b} \sim 1/6$.  Thus, the off-diagonal
coupling to the $Z$ boson is necessarily non-zero.

The viability of the light bottom squark scenario has been questioned 
on the grounds that the heavier $\tilde{b}_2$ should have been detected 
at LEP-II.  The argument is  based on the evaluation of 
SUSY-QCD corrections to the $Zb\bar{b}$ vertex in  
Ref.~\cite{Cao:2001rz} within the context of the light bottom squark and light 
gluino scenario.  These loop corrections contribute negatively to $R_b$, the 
ratio of the width for $Z \rightarrow b \bar{b}$ to the total hadronic 
width, and they increase in magnitude with the mass of $\tilde{b}_2$. 
To maintain consistency with data, the authors of Ref.~\cite{Cao:2001rz} 
argue that the mass of $\tilde{b}_2$ must be less than 125 (195) GeV at the 
$2\sigma$ ($3\sigma$) level. In an extension of this analysis, Cho 
claims that $\tilde{b}_2$ must be lighter than $180$ GeV at
the $5\sigma$ level~\cite{Cho:2002mt}.  A $\tilde{b}_2$ in the mass range 
$< 200$~GeV could have been produced in association with a light 
$\tilde{b}_1$ at LEP-II 
energies, and since no claim of observation has been made, the authors of 
these studies suggest that LEP data disfavor the light bottom squark 
scenario.  A heavier $\tilde{b}_2$ ($\gtrsim 200$ GeV) is allowed if 
$CP$-violating phases are present~\cite{Baek:2002xf}. Real decays such 
as 
$Z \rightarrow \tilde{b}_1 \bar{b} \tilde{g} + \tilde{b}^*_1 b \tilde{g}$ 
contribute positively to $R_b$~\cite{Cheung:2002na} and soften these bounds
to 160 (290) GeV at the $2\sigma$ ($3\sigma$) level~\cite{Luo:2003uw}.  
We remark that experimental searches for SUSY particles are model-dependent 
and a search of LEP-II data for a $\tilde{b}_2$ in the light bottom squark 
scenario has not yet been undertaken.  
The cross sections and discussion of decay 
modes in this paper may help to motivate such a search. 

We begin with the predicted cross sections 
and event rates for production of $\tilde{b}_1 \tilde{b}^*_2$ pairs at 
energies explored at the CERN LEP collider.  For the large heavy bottom 
squark masses that we consider, LEP-II is unable to pair-produce heavy
bottom squarks, and off-diagonal production is the only viable option.
We then discuss decay of 
$\tilde{b}_2$, the heavier of the two bottom squarks.  We 
consider the dominant decay mode $\tilde{b}_2 \rightarrow b \tilde{g}$, 
and we evaluate the total width for this decay as a function of $m_2$, the 
mass of $\tilde{b}_2$.  Subsequently, taking gluino decay into account,  
we present and evaluate the amplitude for the full three-body 
decays $\tilde{b}^*_2 \rightarrow b \bar{b} \tilde{b}^*_1$  and 
$\tilde{b}^*_2 \rightarrow \bar{b} \bar{b} \tilde{b}_1$.  The Majorana 
nature of the gluino permits final states in which there can be bottom 
quarks of the same sign ({\em i.e.}, $b b$ or $\bar{b} \bar{b}$) as well 
as the $b \bar{b}$ configurations expected in SM situations.  The overall 
process, $e^+ e^- \rightarrow \tilde{b}_1 \tilde{b}^*_2$, followed by 
$\tilde{b}_2$ decay leads to a four-parton final state.

\begin{figure}
\includegraphics[height=10.0cm]{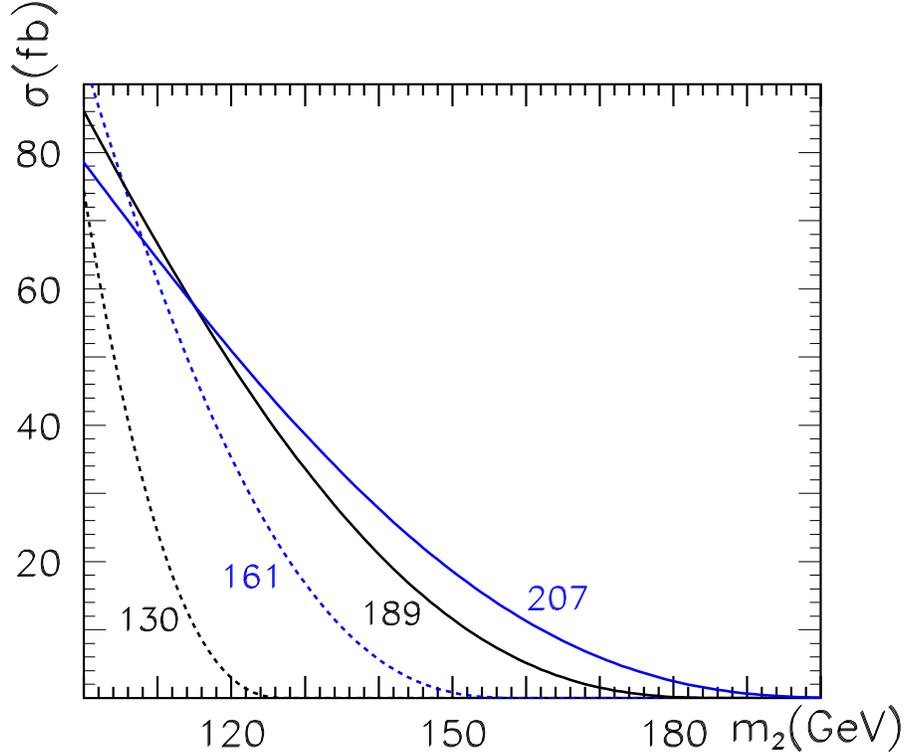}
\caption{Cross section for the process
$e^+ e^- \to \tilde{b}_1 {\tilde{b}}^*_2$ at center-of-mass 
energies 130, 161, 189, and 207 GeV, as a function of the mass 
$m_2$ of the heavier bottom squark. The mass of the lighter 
bottom squark is $m_1 = 3.5$~GeV, and $\sin^2 \theta_b 
= 1/6$.}
\label{fig:lepcrsec}
\end{figure}
\subsection{Cross Sections and Event Rates}

Selecting center-of-mass energies spanning those at which data were 
accumulated at the CERN LEP-II facility, we show the cross section for 
$\tilde{b}_1 {\tilde{b}}^*_2$ production as a function of the 
mass $m_2$ in Fig.~\ref{fig:lepcrsec}.  In this illustrative calculation, 
the mass $m_1$ of the lighter bottom squark is $m_1 = 3.5$~GeV, and 
$\sin^2 \theta_{b} = (2/3)\sin^2 \theta_W \simeq 1/6$.  Focusing 
on the energy dependence at $m_2 = 100$~GeV, we notice that the cross 
section grows with center-of-mass energy $\sqrt{s}$ from 130 to 161 GeV  
and then falls as energy increases.  This behavior may be traced to the 
combined influences of the $|\mathbf{p}^*|^3$ threshold suppression and 
the usual $1/s$ dependence at large $s$.

\begin{figure}
\includegraphics[height=10.0cm]{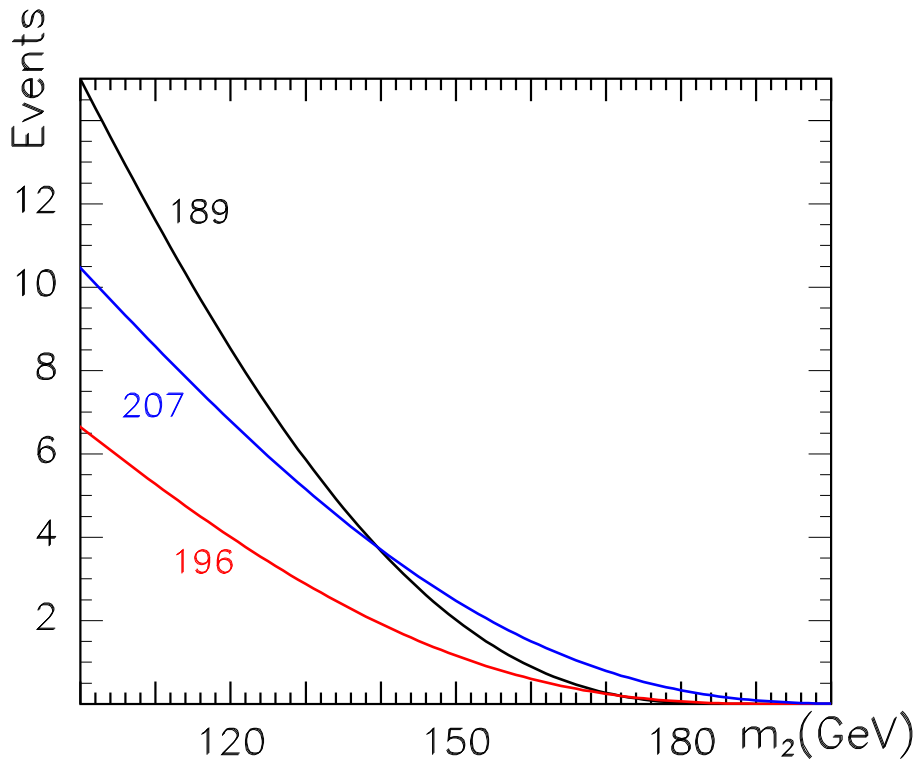}
\caption{Number of events from 
$e^+ e^- \to \tilde{b}_1 {\tilde{b}}^*_2$ at center-of-mass energies 
189, 196, and 207 GeV as a function of the mass $m_2$ of the heavier 
bottom squark. The mass of the lighter bottom squark is $m_1 = 3.5$~GeV, 
and $\sin^2 \theta_b = 1/6$.}
\label{fig:lepevents}
\end{figure}

Multiplying by the accumulated integrated luminosities per 
experiment~\cite{BolekClara} at 
LEP-II, we use our cross sections to compute the predicted number of events 
produced as a function of $m_2$.  These results are shown in 
Fig.~\ref{fig:lepevents}.  Below LEP-II center-of-mass energy 189 GeV, the 
integrated luminosities were too small to have produced an appreciable sample 
of events for the process of interest to us.   In order to translate the 
event rates in Fig.~\ref{fig:lepevents} into limits on the observability of 
$\tilde{b}_2$, we must discuss likely decay modes, experimental 
efficiencies, and backgrounds.  Decays are discussed in the next subsection.
Here we remark simply that if at least 5 events are deemed necessary, 
the raw event rates in Fig.~\ref{fig:lepevents} suggest that a bottom squark 
with mass greater than 140 GeV will have escaped detection at LEP-II.   
We present more rigorous estimates below.

\subsection{$\widetilde{b}_2$ Decay}

In a scenario in which the gluino $\tilde{g}$ is lighter than 
$\tilde{b}_2$, the most likely decay process is 
$\tilde{b}_2 \rightarrow b \tilde{g}$.  
As derived below, both the $\tilde b_2$ and $\tilde{g}$ widths are 
narrow compared to their masses, and thus
the description of the three-body decay 
$\tilde b_2 \rightarrow b b \tilde b_1$ in two sequential steps,
$\tilde b_2 \rightarrow \tilde{g} b$ followed by 
$\tilde{g} \rightarrow b \tilde b_1$ is an accurate one.
The width for $\tilde b_2 \rightarrow \tilde{g} b$, in the $m_b =0$ limit, 
is
\begin{eqnarray}
\Gamma(\tilde{b}_2 \rightarrow \tilde{g} b )&=&
\frac{2 \alpha_s (m_2)}{3} \: m_2 \:
\left( 1 - \frac{m^2_{\tilde{g}}}{m^2_2} \right)^2 .
\label{eq:widths2}
\end{eqnarray}
In this expression, $m_{\tilde{g}}$ denotes the gluino mass.  In the limit 
$m_2 \gg m_b, m_{\tilde{g}}$, 
the width grows linearly with $m_2$, as expected.  
To estimate the magnitude of 
$\Gamma(\tilde{b}_2 \rightarrow \tilde{g} b )$, 
we adopt a gluino mass within the range obtained in the light gluino 
and light bottom squark scenario: 
$12 < m_{\tilde{g}} < 16$~GeV.  
The full width is more than  
an order of magnitude smaller than the mass $m_2$,
as expected since the relative size is controlled by
$\alpha_s (m_2) \sim 0.1$.  For example, choosing 
$m_2 = 150$ GeV and $m_{\tilde{g}} = 15$ GeV, we find 
$\Gamma_{\tilde{b}_2} \simeq 10$ GeV.
In our subsequent treatment of the process 
$e^+ e^- \rightarrow \tilde{b}_1 \tilde{b}^*_2$, with 
$\tilde{b}^*_2  \rightarrow b \bar{b} \tilde{b}^*_1$ or  
$\tilde{b}^*_2  \rightarrow \bar{b} \bar{b} \tilde{b}_1$, 
we are justified in 
adopting the narrow width approximation for $\tilde{b}^*_2$,
factorizing the production and decay.  

In the light gluino and light bottom squark scenario, the gluino decays with 
100\% branching fraction into a bottom quark and and a light bottom squark.  
However, since the gluino is Majorana in nature, it may decay into either a 
bottom quark or a bottom anti-quark:  
$\tilde{g} \rightarrow b \tilde{b}^*_1$ or 
$\tilde{g} \rightarrow \bar{b} \tilde{b}_1$.  As an intermediate step in our 
full calculation of the width for the three-body decay of $\tilde{b}_2$, we 
first evaluate the width for on-shell gluino decay.  
The decay width for the two-body subprocess 
$\tilde{g} \rightarrow b \tilde{b}$ is 
\begin{eqnarray}
\Gamma_{\tilde{g}} \equiv \Gamma(\tilde{g}\to b \tilde{b}_1^*) +  
\Gamma(\tilde{g}\to \bar{b} \tilde{b}_1) &=& 
\frac{\alpha_s (m_{\tilde{g}})}{4} \: m_{\tilde{g}} ,
\end{eqnarray}
where the small bottom and light bottom squark masses are neglected.
For $m_{\tilde{g}} = 15$ GeV, we find $\Gamma_{\tilde{g}} = 0.6$ GeV.  We note
that corrections to this expression from the finite bottom quark and 
bottom squark masses
are generally not negligible, and introduce a dependence on the squark mixing
angle.  However, in all cases $\Gamma_{\tilde{g}} \ll m_{\tilde{g}}$, and
these corrections have little effect on the heavy bottom squark width 
or the 
distributions of the (highly boosted) gluino decay products from
$\tilde{b}_2$ decays.

\begin{table}[t]
\caption{The decay widths 
$\Gamma_{\textrm{LS}}$ and
$\Gamma_{\textrm{OS}}$ in GeV obtained
with $m_{\tilde{g}}$=15 GeV, $\sin^2 \theta_b = 1/6$, 
and $\alpha_s(m_2)=$0.116, 0.112, 0.109 and 0.108 
for $m_2=$100, 125, and 150, and 175 GeV, respectively.  Also shown
is the total $\Gamma_{\tilde{b}_2}$.
}
\begin{ruledtabular}
\begin{tabular}{ccccc}
$m_2=$                 & 100 GeV & 125 GeV & 150 GeV  & 175 GeV  \\ 
$\Gamma_{\textrm{LS}}$ & 3.8 GeV & 4.6 GeV & 5.4 GeV  & 6.2 GeV  \\
$\Gamma_{\textrm{OS}}$ & 3.6 GeV & 4.5 GeV & 5.4 GeV  & 6.2 GeV  \\
$\Gamma_{\tilde{b}_2}$ & 7.4 GeV & 9.1 GeV & 10.8 GeV & 12.4 GeV \\
\end{tabular}
\end{ruledtabular}
\label{tab:widths}
\end{table}

\subsection{$ \widetilde{\bm{b}}$$\bm{{}^*_2}$
            $\to$ $b\bar{b}$ $\widetilde{\bm{b}}$$\bm{{}^*_1}$
 and        $\widetilde{\bm{b}}$$\bm{{}^*_2}$
            $\to$ $\bar{b}\bar{b}$ $\widetilde{\bm{b}}$$\bm{{}_1}$
           }

\begin{figure}
\includegraphics[height=3.0cm]{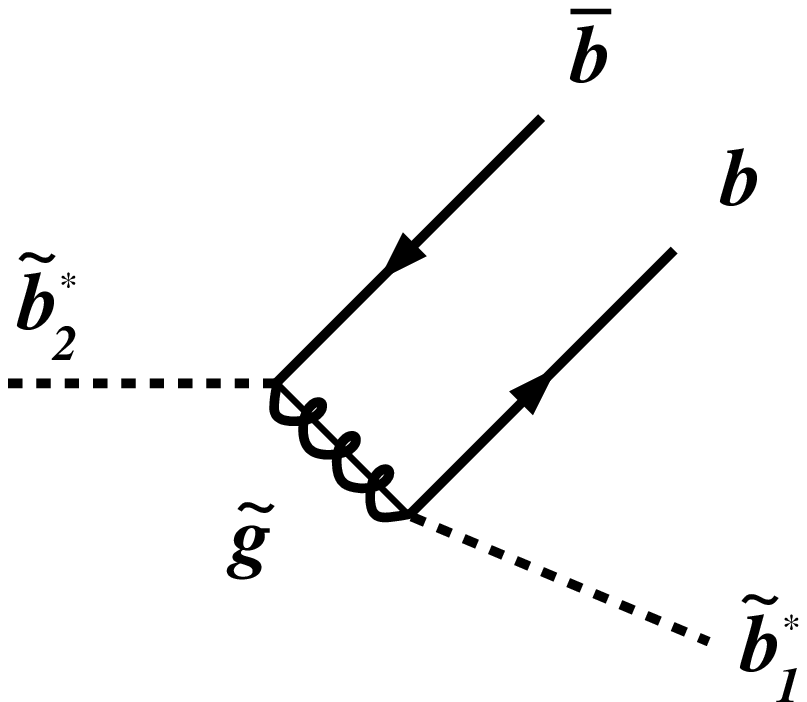}
\caption{Feynman diagram for the decay 
$\tilde{b}^*_2 \to b \bar{b} \tilde{b}^*_1$.}
\label{fig:bbbar}
\end{figure}   

\begin{figure}
\includegraphics[height=3.0cm]{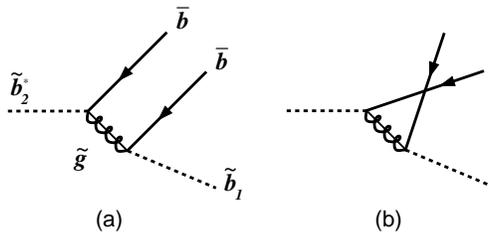}
\caption{Feynman diagrams for the decay 
$\tilde{b}^*_2 \to \bar{b} \bar{b} \tilde{b}_1$.}
\label{fig:bb}
\end{figure} 

In this subsection, we address the full three-body decay 
subprocesses $\tilde{b}^*_2 \to b \bar{b} \tilde{b}^*_1$ and 
$\tilde{b}^*_2 \to \bar{b}\bar{b} \tilde{b}_1$.  The relevant Feynman 
diagram in the case of opposite sign `OS' production ($b \bar{b}$) is 
shown in Fig.~\ref{fig:bbbar}, 
and the two diagrams for like-sign `LS' production in Fig.~\ref{fig:bb}.
For OS and LS decay, our kinematic notation is 
\begin{subequations}
\begin{eqnarray}
\textrm{OS}&:&
\tilde{b}_2^*(s_2)\to 
b(p_1)+\bar{b}(p_2)+\tilde{b}_1^*(p_{\tilde{b}}) ;
\\
\textrm{LS}&:&
\tilde{b}_2^*(s_2)
\to\bar{b}(p_1)+\bar{b}(p_2)+\tilde{b}_1(p_{\tilde{b}}) ,
\end{eqnarray}
\end{subequations}
where the quantities in parenthesis are 4-momenta.
In evaluating the amplitudes for these decays, we must contend with 
the fact that the gluino goes onto its mass-shell within the physical 
region.  To handle this singularity, we resum a class of contributions to
the imaginary part of the
gluino 2-point function to all orders, replacing the gluino propagator 
by a Breit-Wigner form, $p^2 - m^2_{\tilde{g}} \rightarrow 
p^2 - m_{\tilde{g}}^2 + im_{\tilde{g}}\Gamma_{\tilde{g}}$, and we use 
the expressions for $\Gamma_{\tilde{g}}$ above. 
Explicit expressions for the matrix elements and decay widths are 
presented in the Appendix.  
  
Defining
\begin{subequations}
\begin{eqnarray}
\Gamma_{\textrm{LS}}&=&
\Gamma(\tilde{b}^*_2\to \tilde{b}_1+\bar{b} \bar{b}) ,
\\
\Gamma_{\textrm{OS}}&=&
\Gamma(\tilde{b}^*_2\to \tilde{b}_1+b \bar{b}) , 
\end{eqnarray}
\end{subequations}
we provide numerical values of these two widths in Table~\ref{tab:widths} 
for four interesting values of $m_2$, $m_{\tilde{g}} = 15$ GeV, and
$\sin^2 \theta_{b}= 1/6$.  For comparison, we also present the inclusive
$\tilde{b}_2$ width computed from the two-body decay matrix elements,
Eq.~(\ref{eq:widths2}).  It is notable that the LS width is substantial in 
all cases, and is in fact slightly larger than the OS width for the lighter
$\tilde{b}_2$ masses we consider.  The sum of the LS and OS widths, obtained 
from the three-body decay amplitudes, equals to good accuracy the inclusive 
width obtained from the two-body decay process.  In general, the decay widths 
may depend on the sign of the product $\cos\theta_b \sin\theta_b$, but 
this dependence is absent in the limit that $m_b$ and $m_1$ vanish.  

Production of like-sign pairs, attributable directly to the Majorana 
nature of the gluino means that the subprocess of interest here generates  
{\em apparent} ``time = zero'' $B^0 - \bar{B}^0$ flavor-anti-flavor 
mixing in $e^+ e^-$ annihilation at LEP-II and linear collider energies.   
However, the $\tilde{b}_2^* \tilde{b}_1$ production process results in one
or two jets in addition to the jets containing the $b$ and $\bar{b}$, and 
thus these events may not be included in a measurement of 
$B^0 - \bar{B}^0$ mixing that 
focuses on $b \bar{b}$ production without additional radiation. 
The fraction of the decays that lead to like-sign pairs of $b$'s is 
\begin{eqnarray}
B_{\textrm{LS}}&=&
\frac{\Gamma_{\textrm{LS}}}{\Gamma_{\textrm{LS}}+\Gamma_{\textrm{OS}}}
\approx
\frac{\Gamma(\tilde{g}\to \tilde{b}_1+ \bar{b})}
     {\Gamma(\tilde{g}\to \tilde{b}_1+ \bar{b})
     +\Gamma(\tilde{g}\to \tilde{b}^*_1+ b)} ,
\end{eqnarray}
and from Table~\ref{tab:widths}, we see that this ratio is close to 
$\frac{1}{2}$ for all of the heavy bottom squark masses of interest, with 
LS slightly dominant for smaller $\tilde{b}_2$ masses.  

\subsection{Signatures and Discovery Potential at LEP-II}

The overall process, $e^+ e^- \rightarrow \tilde{b}_1 \tilde{b}^*_2$, 
followed by $\tilde{b}_2$ decay leads to a four-parton final state.  
The light bottom squarks carry color and are expected to be observed as 
hadronic jets.  Absent model-dependent assumptions about bottom squark 
decays, these jets may have no special flavor content.  Our SUSY subprocess 
results therefore in a four-jet final state: with 2 $b$ jets and 
2 $\widetilde{b}$ jets. 
 
The massive $\tilde{b}^*_2$ is produced in 
$e^+ e^- \to \tilde{b}_1 \tilde{b}^*_2$ with relatively little  
momentum.  The three products from its decay will therefore inherit 
little sense of direction.  The distribution in $\cos\theta_{1j}$ will 
tend to be fairly flat; subscript $1$ denotes the primary $\tilde{b}_1$,  
and $j$ labels one of the decay products of the $\tilde{b}_2$.
The heavy parent $\tilde{b}^*_2$ tosses off a $\bar{b}$ and a gluino, both 
with substantial but oppositely directed momentum.  The gluino then 
decays into a $\tilde{b}_1$ and the second $b$ or $\bar{b}$, with the 
daughter particles retaining the direction of the gluino's momentum.  
Since the daughter $b$ or $\bar{b}$ follows the direction of the 
gluino, the two final $b$'s, 
in the event, whether like-sign or opposite-sign, emerge in opposite 
hemispheres in the overall $e^+ e^-$ system.  The invariant mass 
of the two final $b$'s will tend to be large.  Furthermore,
since $m_{\tilde{g}} \ll m_2$, we expect the $\tilde{g}$
to be highly boosted, with a small opening angle between its decay products.

The two $b$ jets are predicted to emerge in a 
fairly back-to-back configuration, much as is expected from a standard model 
QCD subprocess $e^+ e^- \rightarrow (\gamma,Z^*) \rightarrow b \bar{b} g$, 
with $g \rightarrow$ one or more ${\rm jet}$s.  (One of the $\tilde{b}$ 
jets may emerge fairly close to one of the $b$ jets, resulting in a three-jet 
topology, as we investigate below.)  The 
configuration produced by the SUSY process differs from that associated with 
$e^+ e^- \rightarrow (\gamma,Z^*) \rightarrow q \bar{q} g$ 
with $g \rightarrow b \bar{b}$.  In this later process, gluon splitting would 
yield $b \bar{b}$ pairs with modest invariant mass.  

The key question is how heavy the second scalar bottom quark 
$\tilde{b}_2$ might be and still
be discovered lurking in the LEP-II data.  Alternately, one can ask what
range of heavy bottom squark masses are ruled out by the data, if no signal is 
observed.  We concentrate on the high luminosity run at $\sqrt{s} = 207$ GeV
at LEP-II, as it provides the greatest number of events for the masses of
interest (c.f. Fig.~\ref{fig:lepevents}).  For this analysis, we do not 
distinguish the LS $\bar{b} \bar{b}$ and OS $b \bar{b}$ situations, adding 
the distributions generated in the two cases.  There would likely be greater 
potential for identifying the SUSY events if there were experimental 
capability to separate $b$ and $\bar{b}$ jets.  

To answer our question, we address first the experimental signature of the 
off-diagonal bottom squark production process. 
The two bottom quark jets are almost always distinguishable, with
a large separation between them.  Similarly, the ``primary'' light bottom 
squark tends to be visible as a distinct jet of hadrons.  However, the 
light bottom squark from the $\tilde g$ decay is often rather collinear 
with the bottom quark from the same decay, because the $\tilde g$ tends 
to be boosted by the heavy bottom squark mass.  Thus, we must establish 
the number of distinctly observable jets in the final state.

We define an observable jet as one with transverse momentum ($p_T$) greater
than 10 GeV, lying in the central region of the detector, $|y| \leq 2$,
where $y$ is the jet rapidity.  The
separation between two jets is quantified by 
$\Delta R \equiv \sqrt{\Delta y^2 + \Delta \phi^2}$, where
$\Delta \phi$ is the difference in azimuthal angles.
We consider two
jets distinct from one another provided $\Delta R \geq 0.4$; for smaller
$\Delta R$ they are merged into a single jet.  The distribution
in the number of jets depends on the $\tilde b_2$ mass.  
For $m_2 \sim 120 $ GeV, we find that
the rate is split roughly evenly between
$3$ jets and $4$ jets.  For a heavier $m_2 \sim 150 $ GeV, the rate 
is split roughly as $2/3$ $3$-jet events and $1/3$ $4$-jet events.  The 
difference arises 
because the larger $\tilde b_2$ mass results in a more highly boosted
gluino, and thus more collinear decay products which are more likely to be
merged into a single jet.  For both masses, the $2$-jet rates are smaller
than a few per cent of the inclusive cross sections.  
We focus on detection of the $4$-jet channel because its rate is a large 
fraction of the total rate for all masses of 
interest, and because we expect that the $4$-jet configuration has smaller 
backgrounds, 

Backgrounds arise from $e^+ e^- \rightarrow 3$ jets and $4$ jets, respectively,
and involve a variety of mixed QCD-electroweak and purely weak processes.  We
simulate the $4$-jet background using matrix elements from 
MADGRAPH~\cite{Murayama:1992gi}.  After the acceptance cuts described 
above, we find backgrounds are typically much larger than the signal 
rates, hundreds of fb
compared to 30 fb (6 fb) for $m_2 = 120$ GeV (150 GeV).  These may
be reduced by positing a mass for the heavy bottom squark, and demanding that 
three of the jets reconstruct this mass within some window.  Since the width 
of the heavy bottom squark is generally of order 10 GeV in 
the mass range of interest, we consider an invariant 
mass cut such that any three
of the jets reconstruct an invariant mass within 10 GeV of $m_2$.
This $m_2$-dependent cut thus forces the background to vary with
the hypothesized value of $m_2$.  After its application, we find
that the background is reduced to the manageable levels of 20 fb (43 fb)
at 120 GeV (150 GeV).

For both masses, the number of combined signal and background events is
$\geq 8$, and thus one may apply Gaussian statistics to determine the
statistical significance in the usual way, with $\sigma \equiv S / \sqrt{S+B}$
providing the confidence level (CL) for an observed signal, with $S$ signal 
events and $B$ background events.  The resulting significances are about $5\sigma$ 
($0.4 \sigma$) at $m_2 = 120$ GeV (150 GeV).  Thus, LEP-II should
be able to discover the heavier bottom squark through off-diagonal production if
its mass is less 120 GeV.  If no signal is observed, we estimate that masses
less than 130 GeV can be excluded at the $95\%$ CL.  Our analysis could be 
improved in a number of ways, notably if experimental acceptances and 
efficiencies were incorporated, a task beyond the scope of this work.  It is our 
hope that the 
analysis in this paper and the exciting prospect of the discovery of SUSY will 
motivate a detailed search for signals in existing LEP-II data.

\section{Conclusions}
\label{sec:conclusions}

Mixing between weak doublets and singlets is a novel feature of the 
scalar fermion sector of the MSSM.  Aside from leading to interesting
phenomenology, this mixing (in the top squark sector) is also important, 
allowing for large radiative corrections to the mass of the 
lightest Higgs boson, needed if the MSSM is to survive the challenge of 
the negative Higgs boson searches at LEP-II.  
Turning this statement around, once scalar tops are observed, and their 
masses and mixing determined, one can, to a fair degree of accuracy, determine 
whether the theory at the TeV scale is described by the MSSM, or by some more
exotic extension.

In this article we explore the off-diagonal squark production mode
at $e^+ e^-$ colliders as a means to measure the mixing angle and 
to learn more about the MSSM itself.  Squark mixing arises from a
combination of supersymmetric interactions and SUSY-breaking masses
and trilinear terms.  These in turn are related closely to the flavor 
structure of the MSSM.  While there are many reasons to prefer the MSSM as
the theory of physics beyond the standard model, its intense flavor 
problem indicates that SUSY-breaking is somehow very special such that
flavor violation has not yet been observed in low energy experiments.  
Measurements of the flavor-full soft parameters will be the first 
experimental indications as to how nature
has chosen to solve the SUSY flavor problem, and thus how SUSY is broken and
the breaking communicated to the MSSM fields.

Squark mixing is also a key player in defining the properties of the squarks
themselves, determining the coupling to the massive electroweak bosons.  Light
bottom squarks, an interesting ingredient in the supersymmetric resolution of
the bottom quark production cross section at hadron 
colliders~\cite{Berger:2000mp}, escape detection at LEP-I because
their mixing angle is such that the left-handed and right-handed interactions
with the $Z$ boson cancel each other.  This feature necessarily implies that 
off-diagonal production is non-zero and can be used to discover or constrain
the mass of the heavier bottom squark.  With a careful, dedicated analysis of 
existing LEP-II data, we show in this paper 
that it should be possible to discover heavy bottom 
squarks at the $5\sigma$ level with masses as large as 120 GeV.  If no
signal is observed, exclusion limits at the $95\%$ CL 
should be feasible for masses of the order of 130 GeV and less.  Off-diagonal 
squark production thus allows one to explore a
large portion of the parameter space of the light bottom squark scenario.

\begin{acknowledgments}

We acknowledge valuable assistance from Jing Jiang and discussions
with A.~Freitas, M.~Schmitt and C.~E.~M.~Wagner. E.~L.~B. acknowledges 
the hospitality of the Aspen Center for Physics while this paper was 
being completed.  
The research of E.~L.~B. and J.~L.~in the High Energy Physics Division
at Argonne National Laboratory is supported by
the U.~S.~Department of Energy, Division of High Energy Physics, under
Contract W-31-109-ENG-38.
Fermilab is operated by Universities Research
Association Inc. under DOE Contract DE-AC02-76CH02000.
\end{acknowledgments}

\appendix

\section*{Appendix}
\label{sec:appendix}

In this Appendix, we present the matrix elements for the full three-body decay 
subprocesses $\tilde{b}^*_2 \to b \bar{b} \tilde{b}^*_1$ and 
$\tilde{b}^*_2 \to \bar{b}\bar{b} \tilde{b}_1$, keeping the dependence on the
two large masses: $m_2$ and $m_{\tilde{g}}$.  The relevant Feynman 
diagram in the case of opposite sign `OS' production ($b \bar{b}$) is 
shown in Fig.~\ref{fig:bbbar}
and the two diagrams for like-sign `LS' production in Fig.~\ref{fig:bb}.
The 4-momenta are labeled as $p_1$ and $p_2$ for the two bottom quarks; in
the case of OS production, $p_1$ refers to the $b$ and $p_2$ to the $\bar{b}$.
The 4-momentum of the light bottom squark is denoted $p_{\tilde{b}}$.
In the LS case, we include in $\overline{|\mathcal{M}|^2}$ 
the symmetry factor $\frac{1}{2}$ for identical particles in the final state.  

The explicit expressions for these invariant amplitudes, 
summed/averaged over final/initial colors and spins are: 

\begin{subequations}
\begin{eqnarray}
\overline{|\mathcal{M}|^2_{\textrm{LS}}} &=&
\frac{4 g_S^4}{3} \bigg\{ \left[
2 \left(1+\cos^2 2 \theta_b \right) 
(p_1 \cdot p_{\tilde{b}})(p_2 \cdot p_{\tilde{b}}) 
+ m_{\tilde{g}}^2 \left( 1 - \cos^2 2 \theta_b \right)
(p_1 \cdot p_2) \right] \left[ |c_1|^2 + |c_2|^2 \right]
 \nonumber \\ & & 
- \frac{2 \sin^2 2 \theta_b}{3}
 \left[ m_{\tilde{g}}^2 (p_1 \cdot p_2) 
+ 2 (p_1 \cdot p_{\tilde{b}}) (p_2 \cdot p_{\tilde{b}}) \right] 
{\rm Re} \: \left( c_1 c_2^* \right)
\bigg\}
\\
\overline{|\mathcal{M}|^2_{\textrm{OS}}} &=&
\frac{8 g_S^4}{3} |c_1|^2 \left\{
\left(1+\cos^2 2 \theta_b \right) m_{\tilde{g}}^2 (p_1 \cdot p_2)
+ 2 \left( 1 - \cos^2 2 \theta_b \right)
(p_1 \cdot p_{\tilde{b}})(p_2 \cdot p_{\tilde{b}})
\right\}
\end{eqnarray}
\end{subequations}
with 
\begin{subequations}
\begin{eqnarray}
c_i & = & \frac{1}{((p_i+p_{\tilde{b}})^2-m_{\tilde{g}}^2)^2 
+ m_{\tilde{g}}^2 \Gamma_{\tilde{g}}^2} \: , \\
\textrm{Re}[c_1 c^*_2]&=&
\frac{((p_1+p_{\tilde{b}})^2-m_{\tilde{g}}^2)
((p_2+p_{\tilde{b}})^2-m_{\tilde{g}}^2) + m_{\tilde{g}}^2 \Gamma_{\tilde{g}}^2}
{\left[((p_1+p_{\tilde{b}})^2-m_{\tilde{g}}^2)^2 
+ m_{\tilde{g}}^2 \Gamma_{\tilde{g}}^2\right]
\left[((p_2+p_{\tilde{b}})^2-m_{\tilde{g}}^2)^2 
+ m_{\tilde{g}}^2 \Gamma_{\tilde{g}}^2\right]} \: .
\end{eqnarray}
\end{subequations}
To obtain the partial widths we integrate these expressions over the three body
phase space for the 3 approximately massless final state particles.  The
resulting partial widths are shown in Table~\ref{tab:widths}.

\end{document}